\documentclass{article}
\usepackage{spconf,amsmath,graphicx}
\usepackage{algorithm,algorithmicx,algpseudocode}
\usepackage{amsmath,amssymb}
\usepackage{multirow}
\usepackage{graphicx}
\usepackage{epstopdf}
\usepackage{hyperref}
\usepackage{color,soul}
\usepackage[caption=false]{subfig}
\usepackage{tabularx}
\usepackage{cite}
\usepackage{amsmath}
\usepackage[english]{babel}
\usepackage{algorithm,algorithmicx,algpseudocode}
\usepackage{enumitem}
\usepackage{breqn}
\usepackage{xcolor}
\usepackage{makecell}
\usepackage{eqparbox}
\usepackage[switch, modulo]{lineno} 
\usepackage{color}
\usepackage{booktabs}
\usepackage{multirow}
\usepackage{soul}
\DeclareMathOperator{\E}{\mathbb{E}}
\newdimen{\algindent}
\algnewcommand\LeftComment[2]{%
\hspace{#1\algindent}$\triangleright$ \eqparbox{COMMENT}{#2} \hfill %
}

\title{A Study of Semi-supervised Speaker Diarization System Using GAN Mixture Model}
\name{Monisankha Pal, Manoj Kumar, Raghuveer Peri, Shrikanth Narayanan}
\address{Signal Analysis and Interpretation Laboratory, University of Southern California, USA}
\begin{document}
\ninept
\maketitle
\begin{abstract}
We propose a new speaker diarization system based on a recently introduced unsupervised clustering technique namely, generative adversarial network mixture model (GANMM). The proposed system uses x-vectors as front-end representation. Spectral embedding is used for dimensionality reduction followed by k-means initialization during GANMM pre-training. GANMM performs unsupervised speaker clustering by efficiently capturing complex data distributions. Experimental results on the AMI meeting corpus show that the proposed semi-supervised diarization system matches or exceeds the performance of competitive baselines. On an evaluation set containing fifty sessions with varying durations, the best achieved average diarization error rate (DER) is 17.11\%, a relative improvement of 33\% over the information bottleneck baseline and comparable to x-vector baseline.
\end{abstract}
\section{Introduction}
\label{sec:intro}

Speaker diarization addresses the problem of determining ``who spoke when" in an audio recording. It has wide ranging applications related to speaker indexing of data \cite{huijbregts2008segmentation}, and forms an integral component of speech \cite{cerva2013speaker} 
and speaker recognition \cite{liu2016investigating} pipelines. A generic speaker diarization system includes (a) a speech activity detection (SAD) module, which separates speech from non-speech parts, (b) speaker segmentation, where the input audio is segmented into speaker homogeneous chunks and enables extraction of speaker discriminative embeddings such as speaker factors \cite{castaldo2008stream}, i-vectors \cite{shum2013unsupervised, senoussaoui2014study}, \emph{x-vectors} \cite{garcia2017speaker, sell2018diarization}, CNN and LSTM based embeddings \cite{cyrta2017speaker, bredin2017tristounet} and d-vectors \cite{wang2018speaker, zhang2018fully} from those audio chunks, and (c) \emph{speaker clustering} that determines the constituent number of speakers in an audio stream and labels each segment with distinct speaker labels (and possibly, identities).
\par
Many recent works on deep neural network based \emph{embedding} extraction \cite{bredin2017tristounet, garcia2017speaker, wang2018speaker} have advanced speaker diarization research with significant performance improvements. They have effectively replaced the previous embedding approaches based on i-vectors for diarization \cite{sell2014speaker, sell2015speaker}. The largely popular x-vector embeddings, which have proven to be more effective than traditional i-vectors especially for short-duration speech, have become the de-facto standard for speaker recognition \cite{snyder2018x} and diarization \cite{garcia2017speaker}. 
\par
However, speaker clustering has been based on mostly unsupervised algorithms over the years. These algorithms include Gaussian mixture model \cite{zajic2017speaker}; agglomerative hierarchical clustering (AHC) based on similarity measures like Bayesian information criterion \cite{cheng2010bic}, generalized log-likelihood ratio \cite{garcia2017speaker}, information bottleneck (IB) \cite{vijayasenan2009information}; mean shift\cite{senoussaoui2014study}; k-means \cite{dimitriadis2017developing}; spectral clustering \cite{wang2018speaker}; integrated linear programming \cite{rouvier2012global} and links \cite{mansfield2018links}. Most recently, few supervised speaker clustering methods such as UIS-RNN \cite{zhang2019fully} and affinity propagation \cite{yin2018neural} have also been proposed for diarization. Despite the success of the above clustering techniques, speaker diarization still remains a challenging task in many real-world applications due to the wide heterogeneity and variability in audio. 
\par
The recent successes of \emph{generative adversarial networks} \cite{goodfellow2014generative} (GANs) in capturing complex data distributions by encoding rich latent structures has attracted a lot of attention. However, it is difficult to train GANs due to the mode collapse problem \cite{metz2016unrolled}. To address this problem, many variants of GAN such as the Wasserstein GAN (WGAN) \cite{arjovsky2017wasserstein}, multi-generator GAN \cite{hoang2017multi}, mixture GAN \cite{arora2017generalization} have been proposed. More recently, the \emph{GAN mixture model} (GANMM), a novel adversarial architecture with a mixture of generators and discriminators, and a classifier trained in an expectation maximization (EM) fashion was introduced \cite{yu2018mixture}. 
This model was shown to be very effective for image and character data clustering \cite{yu2018mixture}. Although the performance of speech recognition and speaker
verification systems has improved dramatically due to the deep learning approaches, most of the existing clustering techniques for speaker diarization are not yet taking full advantage of it. Therefore, it is worthwhile to explore the  potential of neural network based clustering for speaker diarization. In our present work, we adopt and further develop the GANMM framework for audio based speaker clustering. 

\begin{figure*}[!t]
 \centering
 \captionsetup{justification=centering}
  \includegraphics[width=\textwidth]{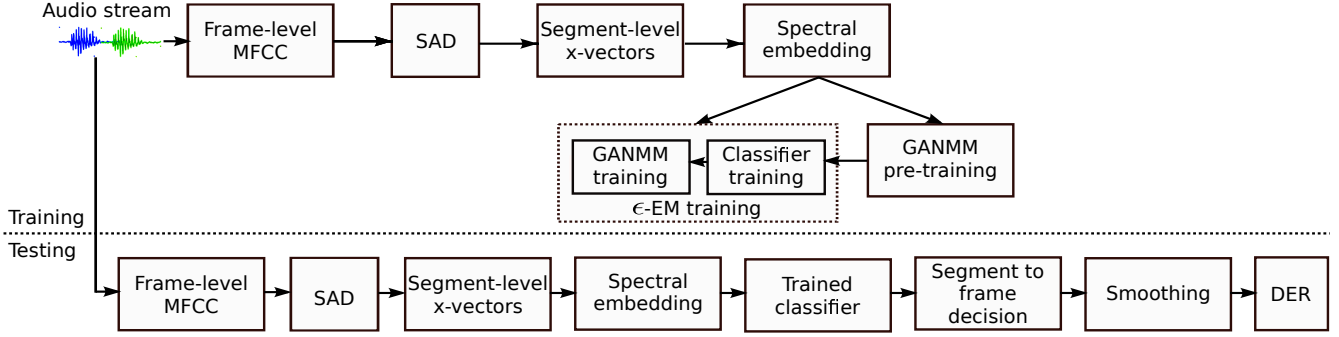}
  \vspace*{-0.5cm}
  \caption{Schematic diagram of the proposed speaker diarization system.}\label{fig1}
  \vspace{-12pt}
\end{figure*}

\par
This work uses x-vector embeddings as a feature representation on short overlapping speech segments. Prior to GANMM clustering, we extract spectral embeddings on the x-vectors to reduce the dimensionality and k-means clustering as initialization for pre-training the GAN models. Training the mixture model is performed through $\epsilon$-EM procedure. The expectation step comprises of learning a classifier to separate the clusters. In the maximization step, each mixture model is trained using the clustered data. Experiments conducted on the AMI corpus confirm the validity of our proposed system. To the best of our knowledge, this is the first attempt of using GANs for speaker clustering in an unsupervised manner within a speaker diarization framework. 

\vspace{-10pt}

\section{Proposed speaker diarization system}

\subsection{Overview of the system}
An overview of the proposed system is presented in Fig. \ref{fig1}. Below, we describe each of the components in detail.

\vspace{-10pt}
\subsection{Segmentation}
In order to isolate potential confounds due to miss and false alarm errors in speech/non-speech detection, we focus solely on the speaker confusion part of diarization error rate (DER). We use oracle SAD for the proposed as well as the baseline systems. After removing the non-speech part in an audio session, x-vectors are extracted from each segment of duration 1.5 sec with 1 sec overlap. While this denser segmentation might help in clustering, each segment may contain more than one speaker. Motivated by the success of spectral clustering in speaker diarization \cite{wang2018speaker}, we exploited the use of spectral clustering on the extracted x-vector embeddings by projecting them into lower-dimensional subspace to produce spectral embeddings using Eigen decomposition.

\vspace{-10pt}
\subsection{Speaker clustering: GANMM training}

We employ GAN models in learning mixture models to capture underlying clusters of complex data. In the following sections, we describe various implementation details involved in GANMM training.
\vspace{-10pt}
\subsubsection{Mitigating early convergence}
Given an audio stream, we assume the extracted embeddings $\mathbf{X} = \{\mathbf{x}_1, \mathbf{x}_2, \ldots , \mathbf{x}_M\}$ to follow probability distribution $p(\mathbf{X})$ and hidden variables $\mathbf{Z}$ with probability distribution $q(\mathbf{Z})$. The log-likelihood with model parameters $\boldsymbol{\theta}$ can be expressed as 
\begin{equation}
    LL(\boldsymbol{\theta}) = LL(q, \boldsymbol{\theta}) + KL(q||p)
\end{equation}
where the log-likelihood $LL(q, \boldsymbol{\theta})$ can be written as $LL(q, \boldsymbol{\theta}) = \sum_{Z} q(\mathbf{Z})\mathrm{ln}(p(\mathbf{X}, \mathbf{Z}|\boldsymbol{\theta})|q(\mathbf{Z}))$ and $KL(q||p)$ is the KL-divergence between $q(.)$ and $p(.)$. In the conventional EM procedure, the E-step matches the hidden variable (clusters in this case) distribution with current posterior probability. The M-step determines the parameters by maximizing the resulted log-likelihood. However, the key issue with GAN models is that 
they can fit the current guess of the hidden variables too well. Hence, the model maximizing the likelihood has extreme value and the whole procedure converges very early with $\boldsymbol{\theta}^{(1)} = \boldsymbol{\theta}^{(2)}$ \cite{yu2018mixture}. 
To mitigate this early convergence, an error term $\epsilon$ is introduced in the E-step with $KL(q^{(t)}||p^{(t)}) = \epsilon_t > 0$. The convergence is guaranteed by keeping $\mathrm{Lim}_{t\rightarrow+\infty}\sum_{i = 0}^{t-1} \epsilon_i < \infty$. In the GANMM setup, an imperfect classifier ($C)$ is trained at the $\epsilon$-E step with generator $(G)$ outputs and the M-step ensures the convergence by eliminating this error. Further details are outlined in \cite{yu2018mixture}.
\vspace{-5pt}
\subsubsection{$\epsilon$-EM procedure}
\begin{figure}[!t]
 \centering
  \includegraphics[width=0.5\textwidth]{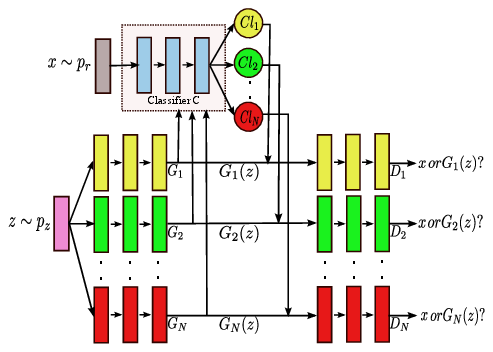}
  \vspace*{-0.6cm}
  \caption{GANMM architecture with $N$ generators, discriminators, and one classifier. Here, $Cl$ represents cluster.}\label{fig2}
  \vspace{-12pt}
\end{figure}

This involves the following:
\\
\textbf{$\epsilon$-E-step:} 
(1) Produce samples from the $N$ generators of the current GANMM model $\boldsymbol{\theta}^{(t)}$ and gather them to construct a data set $S^{(t)} = \{(x^{\prime, (t)}_i, y_i^{(t)})_{i = 1}^{N}\}$, where $x^{\prime, (t)}_i$ is the generated data point from the $i$-th mixture and $y_i^{(t)} = i$ is the generator index. (2) Train an inaccurate classifier $C$ using the data set $S$. (3) Classify each training data of a particular session by the classifier $(C(x_j), j = 1,2, \ldots , M)$ to one of $N$ clusters using maximum probability value.
\\
\textbf{M-step:} Train each mixture of GAN model ($\theta_{i}$) with the clustered data $ D = \{D_{1}, D_{2}, \ldots ,D_{N}\}$ for one generator and several discriminator iterations. Here, as the cluster data is directly fed to the discriminator, discriminator $(D)$ symbol is synonymous for the clustered data. We execute the $\epsilon$-EM step until convergence. Fig. \ref{fig2} illustrates the general architecture of GANMM.
\vspace{-5pt}
\subsubsection{GANMM model}
To this end, the analogy is that each $G$ maps random noise $z$ (sampled from $p_{z}$) to $x = G_i(z)$ and thus induces a single distribution $p_{G_i}$ and, as a whole induces a mixture of $N$ distributions called $p_{model}$ in the data space. Here, one $D$ is supplied for each $G$ by postulating that it will be for one cluster and will distinguish between sample and real training data. 
In contrast, the classifier $C$ performs multi-class classification on training instances $x_j$ into one of the $N$ clusters. The minimax game between these three networks in GANMM can be formulated as
\vspace{-10pt}
\begin{multline}
\scriptsize{\underset{G_{1:N}, C}{\textnormal{min}} \hspace{2pt} \underset{D_{1:N}}{\textnormal{max}} \E_{x \sim p_r} \left[D(x) \right] - \E_{z \sim p_z} \left[\left(D(G(z))\right)\right]}\\\vspace{-10pt} \scriptsize{- \sum_{i = 1}^N \E_{x \sim p_{G_i}}\left[\mathrm{log}C_i(x)\right]}\label{eq1}
\vspace{-10pt}
\end{multline}
\vspace{-2pt}
where the first two terms represent WGAN discriminator and the second term represents WGAN generator cost functions, $C_i(x)$ is the probability that $x$ is generated by $G_i$. The last term in (\ref{eq1}), is a standard cross-entropy loss. It is to be noted that each generator is deemed to produce samples of a specific mode. 

\vspace{-5pt}
\subsubsection{Pre-training}
The rationale for doing pre-training before $\epsilon$-EM training is that if we start $\epsilon$-EM with random parameter initialization, it may result into a bad clustering with all training data is assigned to one cluster. For the pre-training, we prepare data by random shuffling of the training instances and assigning them to clusters. For a better initialization, we first perform k-means clustering with the given number of speakers from ground truth. The GAN model for each cluster is pre-trained for 500 epochs.

\vspace{-5pt}
\subsubsection{Class imbalance problem}
At the beginning of the $\epsilon$-E step, it may be possible that the classifier trained on GAN generated samples assigns clusters with imbalanced instances. This imbalance might get enhanced throughout the iterations with some clusters receiving fewer data than others. To combat this issue, we first ensure that every GAN model generates the same amount of data to train the classifier. Furthermore, for each cluster (say $D_i$), data is augmented from the rest of the clusters (i.e., $D-D_i$) with highest posterior probability to $i$-th cluster according to the classifier. We empirically define the amount of data to be augmented and reduce it along the iterations for convergence.
The full algorithm is presented in Algorithm \ref{algo1}.

\vspace{-5pt}
\subsection{GANMM testing} During inference, the test audio session is uniformly segmented to extract x-vectors, followed by projection onto a lower dimension using spectral embedding. We use the previously trained classifier to predict the cluster decision for each segment which is converted to frame-level. Finally, a median filter with kernel size 361 is applied to smooth the frame-level decisions. 

\vspace{-10pt}

\section{Experimental Evaluation}
\vspace{-5pt}

\begin{algorithm}[!t]
\caption{GAN mixture model clustering algorithm Default values: $\alpha$ = $5\mathrm{e}{-5}$, $m$ = 50, $n_{critic}$ = 5, $\tilde{c}$ = 0.01}
\label{algo1}
\begin{algorithmic}[1]
\Require{$\alpha$: learning rate, $m$: batch size, $N$: number of clusters, $\tilde{c}$: clipping parameter, $N_{epoch}$: number of $\epsilon$-EM iterations, $N_{pt}$: number of pre-training iterations, $\sigma_t$: number of augmented data points at $t$, $n_{critic}$: number of critic iterations for each generator iteration}{}
\State \LeftComment{1} {\textbf{Pre-training}}
\State Do k-means and divide $\mathbf{X}$ into $\{D_1^{(0)}, D_2^{(0)}, \ldots , D_N^{(0)}\}$  
\For{$it$ = 1 to $N_{pt}$}
\For{$i$ = 1 to $N$}
\State $\boldsymbol{\theta}_{D_i}^{(0)}, \boldsymbol{\theta}_{G_i}^{(0)} \longleftarrow \mathrm{trainWGAN} (D_i^{(0)}, n_{critic}, \tilde{c})$
\EndFor
\EndFor
\State \LeftComment{1} {\textbf{$\epsilon$-EM procedure}}
\State t = 0
\For{$it$ = 1 to $N_{epoch}$}
\State t = t + 1
\State \LeftComment{1} {\textbf{$\epsilon$-E-step}}
\State $S^{(t)} = \{(x^{\prime, (t)}_i, y_i^{(t)})_{i = 1}^N\}$ sampled from $N$ generators
\State $g_c \leftarrow \nabla_c\left[\frac{1}{m}\sum_{k = 1}^m \mathrm{crossentropy} \left(C(x^{\prime, (t)}_i), y_i^{(t)}\right)\right]$
\State $\boldsymbol{\theta}_c \leftarrow \boldsymbol{\theta}_c + \alpha. \mathrm{RMSProp}(\boldsymbol{\theta}_c, g_c)$
\State assign $\mathbf{X}$ to $\{D_i\}_{i = 1}^N$ using the classifier $C(\mathbf{X};\boldsymbol{\theta}_c)$
\State \LeftComment{1} {\textbf{M-step}}
\For{$i$ = 1 to $N$}
\State add $\sigma_t$ (reduced along the iterations) instances 
\State from rest of the clusters with highest posterior
\State for cluster $i$ by $C$ to $D_i$.
\State $\boldsymbol{\theta}_{D_i}^{(t)}, \boldsymbol{\theta}_{G_i}^{(t)} \longleftarrow \mathrm{trainWGAN} (D_i^{(t)}, n_{critic}, \tilde{c})$
\EndFor
\EndFor
\end{algorithmic}
\end{algorithm}

\subsection{Data set} We evaluate our proposed algorithm on the popular augmented multiparty interaction (AMI) meeting corpus\footnote{\url{http://groups.inf.ed.ac.uk/ami/download/}}. It consists of about 100 hours of meeting recordings in English and recorded at multiple sites (Edinburgh, Idiap, TNO, Brno). 
For our experiments, we randomly chose ten meetings each of duration 10-20 min as the development set to tune the parameters. Another fifty meetings equally distributed among varying length (10-20 min, 20-30 min, 30-40 min, 40-50 min, 50-60 min) were randomly selected as the evaluation set for benchmarking the proposed system against two baseline diarization systems. Meetings from all the recording sites are present both in development and evaluation set. The proposed system relies solely on each meeting to perform diarization and no separate training data is required.
\vspace{-10pt}

\subsection{Implementation details} 

We use the pre-trained CALLHOME x-vector model\footnote{\url{https://kaldi-asr.org/models/m6}} available in the Kaldi recipe \cite{povey2011kaldi} for x-vector extraction. It is to be noted that in this work unless explicitly mentioned, we have used the ground truth to perform SAD and to calculate the number of speakers in a session, and do the same for both the baselines. 
The discriminator and generator networks in GANMM are feed-forward neural networks with one hidden layer that contains 64 nodes. We use ReLU activation function in the hidden layer and linear activation functions in the input and output layers. The classifier also contains one hidden layer with 64 nodes. 
We use the development data for early stopping during GANMM training and choosing the smoothing window size.
\par
Since our proposed diarization system uses uniform segmentation approach and x-vector as an embedding, we have implemented two methods for diarization as our baselines: Information Bottleneck (IB) \cite{vijayasenan2009information} and x-vector embedding with PLDA scoring and AHC clustering (x-vector) \cite{garcia2017speaker}. To report the performance of different systems, we use the NIST diarization error rate (DER) \cite{fiscus2006rich} performance metric. 

\vspace{-10pt}
\subsection{Results and Discussion}
\vspace{-2pt}
\subsubsection{Results on development set} 
\vspace{-2pt}
We begin with the experiment on the development set for baselines and then applied our basic x-vector based GANMM (P1) to several system refinements (P1-P4) on top of that to arrive at our final proposed system (P4). The experimental results are reported in Table \ref{table1}. Note that for all the results presented in this table, it is assumed that the systems have access to the oracle SAD and number of speakers. From Table \ref{table1}, it is seen that the basic x-vector based GANMM (P1) is not performing well as compared to the baselines. This is due to poor cluster initialization and redundancy in capturing speaker-specific information in high-dimensional embedding space. However, when k-means initialization for pre-training (P2) is introduced in P1, it outperforms IB. We then incorporate spectral embedding only in our P1 system (P3). The significant improvement from P1 to P3 shows that spectral embedding, which retains speaker-specific information more efficiently, is one of the key components for our diarization system. Finally, by employing both spectral embedding on the extracted x-vectors and k-means initialization for GANMM pre-training (P4), we obtain further improvement in performance. 
It is worthwhile to mention that unlike the x-vector baseline, no supervised PLDA was used in any of our experiments. We use our best performing system (P4) for the remaining experiments in this work.

\begin{table}[!t]
\caption{Avg. DER results for the proposed and two baselines on the development set.} 
\centering
\setlength{\tabcolsep}{3pt}
\begin{tabular}{c|cccccc}
\Xhline{2.5\arrayrulewidth}
\multirow{2}{*}{}                      & \multicolumn{6}{c}{System}                            \\ \cline{2-7} 
                                       & IB & x-vector & P1    & P2    & P3   &P4         \\ \Xhline{2.5\arrayrulewidth}
\multicolumn{1}{c|}{Avg. DER (in \%)} & 29.17     & 23.03     & 36.59 & 27.30 & 19.70 & \textbf{18.90} \\ \Xhline{2.5\arrayrulewidth}

\end{tabular}\label{table1}
\vspace{-15pt}
\end{table}

\vspace{-7pt}
\subsubsection{Results on evaluation set} 

Diarization performances in terms of mean and standard deviation in DER on the evaluation set are presented in Table \ref{table2}. We show results by assuming oracle SAD with known number of speakers and estimated number of speakers in column 2 and 3, respectively. For a fair comparison between the proposed and x-vector systems, we implemented a separate diarization system: spectral embedding on the extracted x-vectors with cosine similarity scoring and AHC clustering (denoted as x-vector$^{+}$). 
From Table II, the smaller gain in avg. DER for x-vector system as compared to the proposed is attributed probably due to extra supervised PLDA steps on the x-vectors. On the other hand, spectral embedding on x-vectors brings performance improvement for x-vector$^{+}$ system. 
However, DER improvement between P4 and either of x-vector systems is not statistically significant ($p$ $>$ 0.05 by t-test) 
Moreover, the proposed method provides superior performance over IB baseline. From Table \ref{table2} column 3, we observe that diarization performance degrades for all the systems when estimated number of speakers is used for clustering. In this paper, the number of speakers in a particular session is determined by using eigen gap analysis on the affinity matrix constructed based on cosine similarity between segment x-vectors \cite{wang2018speaker}. However, for x-vector and x-vector$^{+}$ systems, we use thresholding on the PLDA scores to perform AHC clustering for unknown number of speakers. We noticed that, performance degradation is due to mostly under-estimation of the number of speakers for some sessions. We achieve comparable performance for our proposed system as compared to x-vector and x-vector$^{+}$, and significantly better than IB.
\vspace{-10pt}
\subsubsection{Further analysis}

We next check the effectiveness of our proposed system in a variety of practical conditions. Average DER of all the audio files split according to session duration are shown in Table \ref{table3}. For 20-30 min and 30-40 min audio sessions, the proposed system performs better than x-vector and degrades for rest of the audio files. Therefore, we can say that clustering with GANMM results in a comparable performance with supervised methods for the short duration sessions.
\par
For in-depth analysis and to check the effectiveness of our proposed method in more challenging practical scenarios, we first chose meetings from the evaluation set that have majority number of small duration ($<=$ 2.5 sec and 3 sec) speech segments. We report mean DER of the selected sessions in Table \ref{table4} column 2. It is clear from the table that proposed system yields competitive performances to the x-vector baseline system for both cases when fraction of small duration segments within a session is $>=$ 70\%. Both the proposed and x-vector systems are robust to short speech segments.
\par
To show the effectiveness of proposed diarization system in minority speaker detection, we first select meetings from the evaluation set that has at least one speaker who speaks for less than 10\% of the whole meeting duration. 
We then compute the \textit{minority speaker error}, which we define as the fraction of speaker error (in seconds) within the speech from minority speaker over the total session duration. We report the average minority speaker error (in \%) over all sessions in Table \ref{table4} column 3.
It is evident from the table that our proposed system is slightly more robust as compared to the x-vector baseline in non-dominant speaker detection.  
\par
Finally, to evaluate the diarization performance of the proposed GANMM-based system in a scenario with a larger number of speakers, we chose the ICSI meeting corpora \cite{janin2003icsi}. 
The results are shown in Table \ref{table4} column 4. For small number (3-5) of speakers, the proposed system is comparable to the x-vector system; however, its performance deteriorates for a larger number of speakers. Further analysis is required before the proposed diarization system is used for applications with a large number of speakers.  
\vspace{-10pt}

\begin{table}[!t]
\caption{Results on evaluation set for the baseline systems and final version of proposed system.}
\centering
\setlength{\tabcolsep}{3pt}
\begin{tabular}{ccc}
\Xhline{2.5\arrayrulewidth}
System  & \begin{tabular}[c]{@{}c@{}}Mean, std. dev. DER (in \%)\\ (Oracle SAD, known\end{tabular} & \begin{tabular}[c]{@{}c@{}}Mean, std. dev. DER (in \%)\\ (Oracle SAD, estimated \end{tabular} \\ & \#speakers) & \#speakers) \\\Xhline{2.5\arrayrulewidth}
IB              & 25.43 $\pm$ 15.66 &  26.57 $\pm$ 16.04\\ \hline
x-vector        & 15.91 $\pm$ \textbf{7.88}  & 20.14 $\pm$ \textbf{11.36} \\ \hline
x-vector$^{+}$  & \textbf{15.62} $\pm$ 11.41  & \textbf{20.02} $\pm$ 12.98 \\ \hline
P4              & 17.11 $\pm$ 10.57  & 21.56 $\pm$ 12.28  \\ \Xhline{2.5\arrayrulewidth}
\end{tabular}\label{table2}
\vspace{-10pt}
\end{table}

\begin{table}[!t]
\caption{Performance (DER, \%) of x-vector and proposed system on evaluation set split according to session duration.}
\centering
\setlength{\tabcolsep}{3pt}
\begin{tabular}{cccccc}
\Xhline{2.5\arrayrulewidth}
System    & 10-20 min      & 20-30 min      & 30-40 min      & 40-50 min      & 50-60 min      \\ \Xhline{2.5\arrayrulewidth}
x-vector & \textbf{16.54} & 20.33          & 17.57          & \textbf{10.68} & \textbf{14.05} \\ \hline
P4        & 18.39          & \textbf{17.82} & \textbf{12.52} & 18.77          & 18.12          \\ \Xhline{2.5\arrayrulewidth}
\end{tabular}\label{table3}
\vspace{-10pt}
\end{table}

\begin{table}[!t]
\setlength{\tabcolsep}{1pt}
\caption{Performance analysis of x-vector and proposed system in challenging scenarios.}
\centering
\begin{tabular}{ccccccc}
\Xhline{2.5\arrayrulewidth}
\multirow{2}{*}{System} & \multicolumn{2}{c}{\begin{tabular}[c]{@{}c@{}}Avg. DER (in \%) for \\ small \\ speech segments\end{tabular}} & \multicolumn{2}{c}{\multirow{2}{*}{\begin{tabular}[c]{@{}c@{}} Avg. speaker \\ error \\ (in \%)\end{tabular}}} & \multicolumn{2}{c}{\begin{tabular}[c]{@{}c@{}}Avg. DER (in \%) across \\ \#speakers\end{tabular}}                              \\ \cline{2-3} \cline{6-7} 
                        & $\leq$ 2.5 sec                                  & $\leq$ 3 sec                                 & \multicolumn{2}{c}{}                                                                                  & \begin{tabular}[c]{@{}c@{}}\#speakers=\\ 3, 4 and 5\end{tabular} & \begin{tabular}[c]{@{}c@{}}\#speakers=\\ 8, 9 and 10\end{tabular} \\ \Xhline{2.5\arrayrulewidth}
x-vector                & \textbf{15.92}                                                & \textbf{18.84}                                                 & \multicolumn{2}{c}{1.74}                                                                         & \textbf{24.80}                                                         & \textbf{29.28}                                                             \\ \hline
P4                      & 17.61                                                 & 18.92                                                 & \multicolumn{2}{c}{\textbf{1.51}}                                                                         & 26.38                                                         & 34.02                                                             \\ \Xhline{2.5\arrayrulewidth}
\end{tabular}\label{table4}
\vspace{-5pt}
\end{table}


\section{Conclusions}
In this work, we propose a GAN mixture model, a novel deep generative model for speaker clustering within the speaker diarization framework. While the basic x-vector based GANMM is shown to perform poorly, substantial improvement is observed after employing k-means initialization based GANMM pre-training and spectral embedding on the extracted x-vectors. The proposed system results in a relative 33\% DER improvement over the IB baseline and favorably comparable performance to x-vector baseline. 
Furthermore, to the best of our knowledge, this is one of the first approaches exploring the use of GAN mixture model for speaker clustering in the context of speaker diarization. In addition, the proposed diarization system exhibits promising performances in several challenging practical conditions. Future work could investigate variants of GANs in the mixture model in a multi-tasking fashion to further improve diarization performance.

\vfill\pagebreak

\vspace{-13pt}
\bibliographystyle{IEEEbib}
{\bibliography{ref1}}


\end{document}